\documentclass[journal]{IEEEtran}
\ifCLASSINFOpdf
\else
\fi

\usepackage{tikz}
\usepackage{amsmath}
\usepackage{subfigure}
\usepackage{amsthm}

\usepackage{tabularx}
\usepackage{pifont}
\usepackage{filecontents}
\usepackage{amsmath,amssymb,amsfonts}
\usepackage{url}
\usepackage{caption}
\usepackage{colortbl}
\usepackage{flushend}

\definecolor{seagreen}{rgb}{0.18, 0.55, 0.34}
\definecolor{royalpurple}{rgb}{0.47,0.32,0.66}
\definecolor{brown(traditional)}{rgb}{0.59, 0.29, 0.0}
\definecolor{blue}{rgb}{0.3, 0.2, 0.9}
\usepackage[colorlinks,
            linkcolor=blue,
            anchorcolor=blue,
            citecolor=blue]{hyperref}

\begin{document}
%
\title{Generative AI in Data Center Networking: Fundamentals, Perspectives, and Case Study}

\author{Yinqiu~Liu,
        Hongyang~Du,
        Dusit~Niyato,~\IEEEmembership{Fellow,~IEEE},\\
        Jiawen~Kang,
        Zehui~Xiong,
        Yonggang~Wen,~\IEEEmembership{Fellow,~IEEE},
        and
        Dong In~Kim,~\IEEEmembership{Fellow,~IEEE}
        \thanks{Y. Liu, D. Niyato, and Y. Wen are with the College of Computing and Data Science, Nanyang Technological University, Singapore (e-mail: yinqiu001@e.ntu.edu.sg, dniyato@ntu.edu.sg, and ygwen@ntu.edu.sg).}
        \thanks{H. Du is with the Department of Electrical and Electronic Engineering, University of Hong Kong, Pok Fu Lam, Hong Kong (e-mail: duhy@eee.hku.hk).}
        \thanks{J. Kang is with the School of Automation, Guangdong University of Technology, China (e-mail: kavinkang@gdut.edu.cn).}
        \thanks{Z. Xiong is with the Pillar of Information Systems Technology and Design, Singapore University of Technology and Design, Singapore (e-mail: zehui\_xiong@sutd.edu.sg).}
        \thanks{D. Kim is with the College of Information and Communication Engineering, Sungkyunkwan University, South Korea (Email: dongin@skku.edu).
        }
    }
\maketitle
\vspace{-8em}

\begin{abstract}
Generative AI (GenAI), exemplified by Large Language Models (LLMs) such as OpenAI's ChatGPT, is revolutionizing various fields.
Central to this transformation is Data Center Networking (DCN), which not only provides the computational power necessary for GenAI training and inference but also delivers GenAI-driven services to users.
This article examines an interplay between GenAI and DCNs, highlighting their symbiotic relationship and mutual advancements.
We begin by reviewing current challenges within DCNs and discuss how GenAI contributes to enhancing DCN capabilities through innovations, such as data augmentation, process automation, and domain transfer. 
We then focus on analyzing the distinctive characteristics of GenAI workloads on DCNs, gaining insights that catalyze the evolution of DCNs to more effectively support GenAI and LLMs. 
Moreover, to illustrate the seamless integration of GenAI with DCNs, we present a case study on full-lifecycle DCN digital twins. 
In this study, we employ LLMs equipped with Retrieval Augmented Generation (RAG) to formulate optimization problems for DCNs and adopt Diffusion-Deep Reinforcement Learning (DRL) for optimizing the RAG knowledge placement strategy.
This approach not only demonstrates the application of advanced GenAI methods within DCNs but also positions the digital twin as a pivotal GenAI service operating on DCNs. 
We anticipate that this article can promote further research into enhancing the virtuous interaction between GenAI and DCNs.
\end{abstract}


%
\IEEEpeerreviewmaketitle

\section{Introduction}
From 2022, the transformative power of Generative AI (GenAI) has been demonstrated by the phenomenal success of Large Language Models (LLMs) like OpenAI's ChatGPT. 
Unlike Discriminative AI (DAI) models that learn the boundaries among classes, GenAI models are adept at representing the distribution of sample data, enabling them to produce realistic digital content \cite{10515203}.
Nonetheless, to effectively capture complex patterns of real-world samples, GenAI models generally feature complicated architectures with huge sizes. 
Consequently, GenAI operations demand significant computational resources, necessitating robust infrastructural support.

Simultaneously, the field of Data Center Networking (DCN) has undergone substantial development to meet the burgeoning requirements of advanced applications represented by GenAI. 
For instance, NVIDIA presented the Blackwell DCN architecture oriented to LLMs, with optimized tensor cores to accelerate Transformer and 130TB/s switches to accelerate data synchronization during the GenAI training\footnote{https://www.nvidia.com/en-us/data-center}.
Moreover, AI advancements enable DCNs to transcend traditional roles as mere data conduits, efficiently orchestrating data storage, transportation, and computing \cite{9099505}.  
For example, Huawei presented iManager\footnote{https://carrier.huawei.com/en/products/fixed-network}, which utilizes AI models to manage DCN power allocation and is estimated to improve the resource utilization rate of its facilities by 20\%.

The interplay between GenAI/LLMs and DCNs is multifaceted and deeply symbiotic, as advancements in one field drive progress and enable new capabilities in the other.
\begin{itemize}
\item \textbf{GenAI Enhances DCNs}: GenAI significantly outperforms DAI in enhancing DCN capabilities, owing to its inherent capability in distribution representation. For instance, in load balancing \cite{8834843}, GenAI can enrich the training dataset and simulate diverse DCN states. Additionally, GenAI can make accurate predictions in dynamic environments by learning from data distributions. It can even adaptively transfer and fine-tune knowledge across different domains. These advanced capabilities are not available with DAI.
\item \textbf{DCNs Supports GenAI}: DCNs are the cradle of GenAI. Taking ChatGPT as an example, its training was conducted on a super DCN comprising tens of thousands of servers. To meet the escalating demands from GenAI, future DCNs should be specifically designed to accommodate the unique characteristics of GenAI workloads \cite{NSDI}. For instance, considering that the pre-training stage of GenAI is resource-intensive and time-consuming, the load balancing mechanism should be optimized to efficiently distribute tasks across servers, thereby reducing the service latency during high-demand periods.
\end{itemize}
\begin{figure*}[htbp!]
  \centering
  \includegraphics[width=0.93\textwidth]{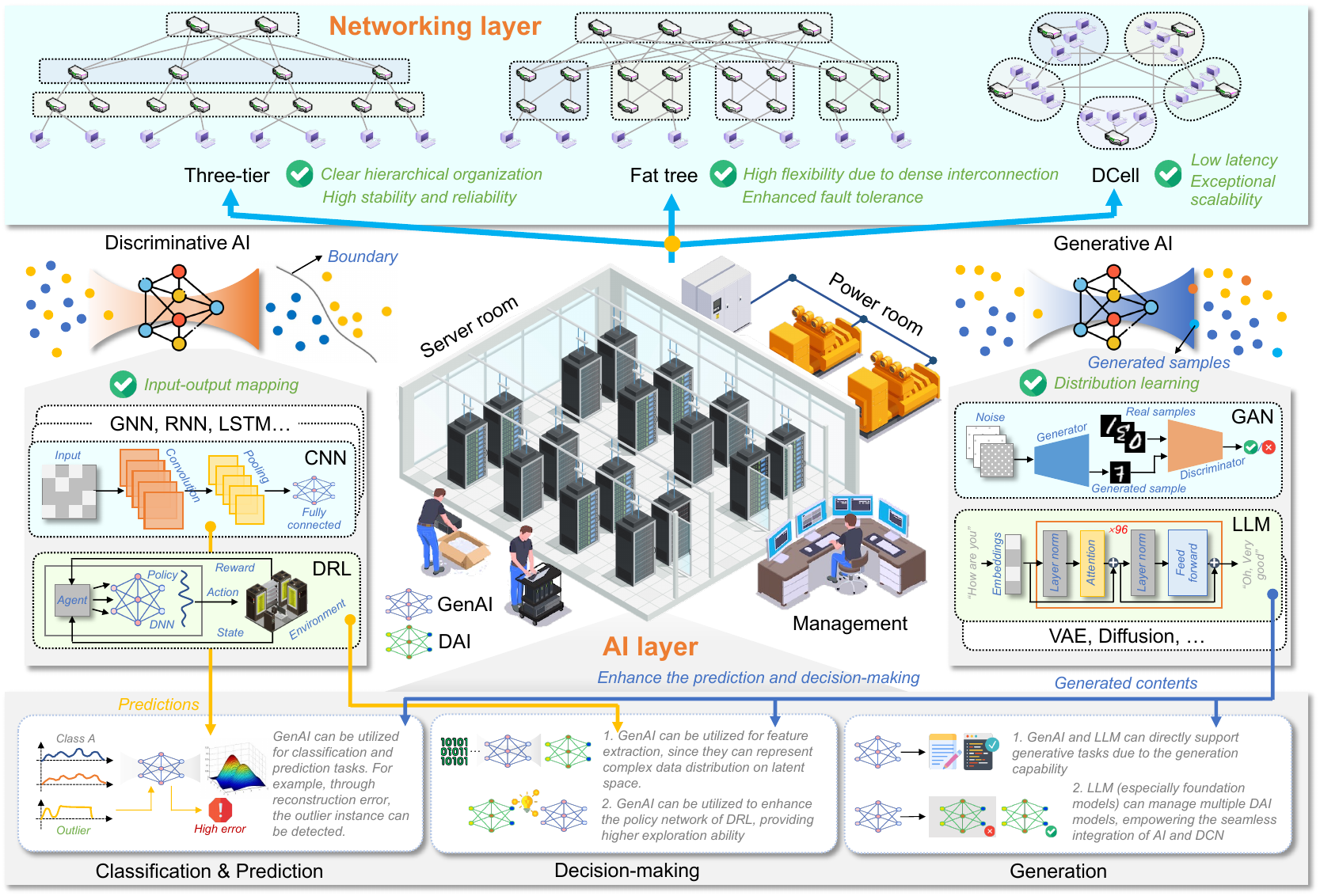} 
  \caption{Illustration of DAI and GenAI applications in DCN. The middle part illustrates the representative data center layout, with the server room, management center, and power room. Numerous servers can be connected in the three-tier, fat-tree, or DCell manners, the so-called networking layer. Finally, DAI and GenAI in the AI layer collaborate to realize efficient and secure DCN operations.} 
  \vspace{-0.4cm}
  \label{DCN}
\end{figure*}

We notice that some preliminary work regarding this topic has been done.
For instance, Li \textit{et al.} \cite{10.1145/3600100.3623719} presented ChatTwin, which leverages GPT-4 to automatically configure DCNs according to the operator's functional requirements.
Hu \textit{et al.} \cite{NSDI} traced the LLM running on DCNs for six months, characterizing the LLM workload and presenting a series of mechanisms to ensure efficient LLM training.
Nonetheless, an interplay between GenAI and DCN has not yet been thoroughly explored in the academic realm, with several unsolved questions, e.g., ``\textit{which DCN tasks can be effectively solved by GenAI and why}" and ``\textit{how DCNs evolve to efficiently support GenAI/LLM workloads}". 
This paper first reviews the DAI-based methods to solve DCN issues, as well as the challenges.
Then, we analyze new possibilities brought by GenAI and review the existing works leveraging GenAI/LLM to enhance DCNs.
In addition to how GenAI enhances DCNs, we intend to explore how DCNs evolve to better support GenAI.
Hence, we analyze the characterization of GenAI/LLM workloads on DCNs.
Based on the acquired insights, we further review the DCN designs catering to GenAI/LLM.
Finally, we perform a case study to showcase the virtuous interaction between GenAI and DCNs.
The main contributions of our paper can be summarized as follows.
\begin{itemize}
    \item \textit{To the best of our knowledge, this is the first work exploring the interplay between GenAI and DCNs.} First, we analyze the existing challenges of DAI-empowered DCN and discuss why and how GenAI can bring new possibilities. Moreover, we review the existing works of GenAI for optimizing DCNs, including data augmentation, anomaly detection, and process automation.
    \item From the opposite perspective, we characterize the pattern of GenAI/LLM workload on DCNs. Based on the insights from characterizations, we review the mechanisms that have been proposed to optimize the running of GenAI/LLM on DCNs. Then, we look forward to the development direction of DCNs in the GenAI era.
    \item We present a case study on full-lifecycle DCN digital twins. The proposal can automatically optimize DCNs via LLM with Retrieval Augmented Generation (RAG) and solve optimization problems via diffusion-Deep Reinforcement Learning (DRL). We evaluate our proposal by minimizing the retrieval latencies. Moreover, our digital twin is hosted by DCNs, revealing the virtuous interaction between GenAI and DCNs.
\end{itemize}

\section{Preliminaries: Data Center Networking, Discriminative AI, and Generative AI}

\subsection{Basics of Data Center Networking (DCN)}
DCN refers to the integration of hardware and software infrastructure that facilitates the interconnection and management of numerous heterogeneous servers within a data center to realize efficient communication and high-quality services \cite{8316818}.
This setup hinges on three essential components, i.e., hardware, software, and protocols.
Specifically, hardware involves various networking devices, such as switches, routers, and firewalls.
Software is deployed on the DCN hardware, realizing efficient network management, protection, and service provisioning.
Finally, protocols define the standards for heterogeneous DCN hardware/software regarding data transportation, storage, etc.
As illustrated in Fig. \ref{DCN}, configuring these components in different ways leads to diverse DCN architectures, such as three-tier, fat tree, and DCell \cite{8316818}.

\subsection{Issues of DCN and Solutions from DAI}
To manage large-scale DCNs that accommodate dense devices, a series of mechanisms should be carefully designed and developed.
In this part, we review the major technical issues of DCNs, as well as the solutions based on DAI models.

\subsubsection{Topology and Routing}
The topology of DCNs indicates how servers are interconnected, impacting the latency and reliability of data transmission. 
Then, efficient routing protocols can be deployed to find the optimal paths for data packets to travel from source to destination. 
Optimizing topology and routing is crucial for minimizing latency, managing bandwidth more effectively, and ensuring resilience against failures.

By harnessing DAI, network operators can precisely model and forecast network behaviors under various load conditions, enabling them to proactively adjust topologies and enhance the overall network performance. For instance, Xie \textit{et al.} \cite{8382151} presented Topology2Vec, which first encodes topological features by graphs and trains a DAI model to optimize controller placement in complicated DCNs.

\subsubsection{Load Balancing}
Load balancing aims to distribute the workload across numerous DCN servers to ensure optimal resource utilization and prevent the impacts of stragglers. Moreover, available resources should be placed and scheduled in a flexible way to fit the workload pattern.
DAI, especially DRL, significantly revolutionizes load balancing in DCNs. Specifically, the efficacy corresponding to each allocation scheme can be modeled by rewards. Afterward, the optimal scheme can be gradually approached through a Markov process, following the DRL paradigm. For instance, Liu \textit{et al.} \cite{8834843} leveraged Deep Q-learning (DQN) to optimize data placements in DCNs and minimize the data movement latency.
 
\subsubsection{Traffic Management}
Traffic management in DCNs involves predicting, controlling, and optimizing both east-west and north-south data flows. Effective traffic management facilitates DCNs to avoid congestion, reduce latency, and improve Quality of Service (QoS). Considering the complex temporal-spatial dependencies existing in network traffic, researchers utilize diverse Deep Neural Networks (DNNs) to infer the relationship between network factors and traffic performance, thus supporting accurate traffic pattern predictions \cite{9272676}. 


\subsubsection{Fault Tolerance and Resiliency}
Resiliency in DCNs is crucial for maintaining high availability and continuous service delivery. 
It involves mechanisms such as anomaly detection and predictive maintenance, which predict and prevent equipment failures before they occur. 
Additionally, disaster recovery protocols allowing rapid response and restoration in the event of a failure are required.
DAI is widely adopted to predict anomalies and failures. 
For instance, Ilager et al. \cite{9272657} trained a multiple Multi-Layer Perception (MLP) model to predict the temperature of DCN servers under different workload conditions, thereby facilitating proactive maintenance.


\textbf{Perspectives}: The above discussion demonstrates that DAI is usually applied in DCNs in two ways.
First, by inferring the relationship between inputs and outputs, DAI exhibits strong predictive ability, enabling it to forecast workload, traffic patterns, and even potential failures accurately. 
Second, DAI, especially DRL, excels in complex decision-making processes, thus greatly facilitating tasks such as workload allocation, routing optimization, and the development of advanced defense mechanisms.

\subsection{Existing Challenges}
Despite the above progress, several challenges persist where DAI cannot effectively aid DCNs. Alternatively, GenAI can potentially offer solutions:
\begin{itemize}
    \item \textbf{Insufficient Training Data}: Given the topological complexity of DCNs, DNNs require substantial scale to represent and learn complicated environmental features. Training large DNNs necessitates vast amounts of precisely labeled data. Nonetheless, there are often challenges in ensuring sufficient training data for DCNs. First, the high cost of labeling massive DCN data, coupled with the rapid changes in network conditions, makes it difficult to gather robust datasets. Additionally, the security-sensitive nature of data in DCNs can restrict the availability of real-world data for training purposes.
    \item \textbf{Limited Versatility}: The applications of DAI models are limited to making predictions/decisions. Specifically, each DAI model is dedicated to performing a specific function. This one-to-one paradigm limits the seamless integration of AI technology in DCNs. Moreover, numerous advanced DCN tasks are generative rather than predictive, such as process automation and responding to unseen issues, which are beyond DAI's capabilities.
    \item \textbf{Weak Adaptability and Generalization}: Another major challenge is regarding the flexibility of DAI models across heterogeneous DCN environments. Each data center has unique configurations, traffic patterns, and security requirements. DAI models trained on one set of network conditions may not perform well in other settings without extensive customization and tuning, making it difficult to adapt to diverse infrastructures. Even though some techniques, such as transfer learning, can help, they are not versatile, complex to employ, or require resource-intensive re-training.
\end{itemize}

\subsection{GenAI and LLM: New Opportunities}
\subsubsection{Learning Objective}
Recall that DAI is trained to make predictions and classifications.
As shown in Fig. \ref{DCN}, for inputs $X$ and outputs $Y$, the learning objective of DAI is to acquire the conditional probability $P(Y|X)$. 
In contrast, GenAI models aim to obtain the joint probability $P(X, Y)$. 
Such a learning objective enables GenAI models to capture intricate details about the underlying input distribution and latent features.
In contrast, the objective of DAI is typically more focused and direct, i.e., mapping input features directly to outputs.
Consequently, DAI is optimal for tasks such as classification but may not capture the broader underlying data characteristics that are not directly relevant to the specific task.

\subsubsection{Model Architecture}
Notable GenAI models include Variational Autoencoders (VAEs), Generative Adversarial Networks (GANs), diffusion models, and transformers \cite{10515203}. 
Specifically, VAEs operate through an encoder-decoder structure. The encoder compresses input data into a latent space representation, capturing essential data features, while the decoder leans to reconstruct the data back to its original form.
Afterward, new instances can be generated by sampling and decoding different points from the latent space \cite{10515203}.
Likewise, GANs consist of two competing models: a generator that creates realistic data and a discriminator that evaluates whether the output is synthetic. 
This architecture trains through a zero-sum game where the generator improves its ability to produce increasingly realistic outputs as the discriminator becomes better at distinguishing real from synthetic \cite{10515203}. 
In contrast, inspired by non-equilibrium thermodynamics, diffusion models operate by a forward diffusion process that gradually disturbs data samples into random noise, during which a denoising network is trained. 
Then, a denoising process is performed to generate new instances that mirror the original distribution through learned noise reduction \cite{10515203}. 

LLMs, exemplified by the Generative Pre-trained Transformer (GPT) \cite{10515203}, are a subset of GenAI specifically designed for understanding and generating natural language.
As such, LLMs can accomplish tasks such as summarization, translation, and question-answering. 
Central to their architecture is the transformer model, which utilizes self-attention mechanisms to process sequential data, such as text. 
Moreover, to enhance the capability of understanding long inputs and generating complicated answers, LLMs typically incorporate numerous transformer layers. 
For instance, GPT-3 accommodates 96 transformer layers, each of which contains 96 attention heads, resulting in a large mode size with 175 billion parameters\footnote{Data available on: https://lambdalabs.com/blog/demystifying-gpt-3}. 
Empowered by ever-enlarging models, alongside technical advancements such as cross-modal attention and mixture-of-experts architectures, LLMs are extending their capabilities beyond text generation. 
Nowadays, multimodal LLMs, such as GPT-4, can effectively understand and generate images, videos, and even 3D objects. 
\begin{figure*}[htbp!]
  \centering
  \includegraphics[width=0.98\textwidth]{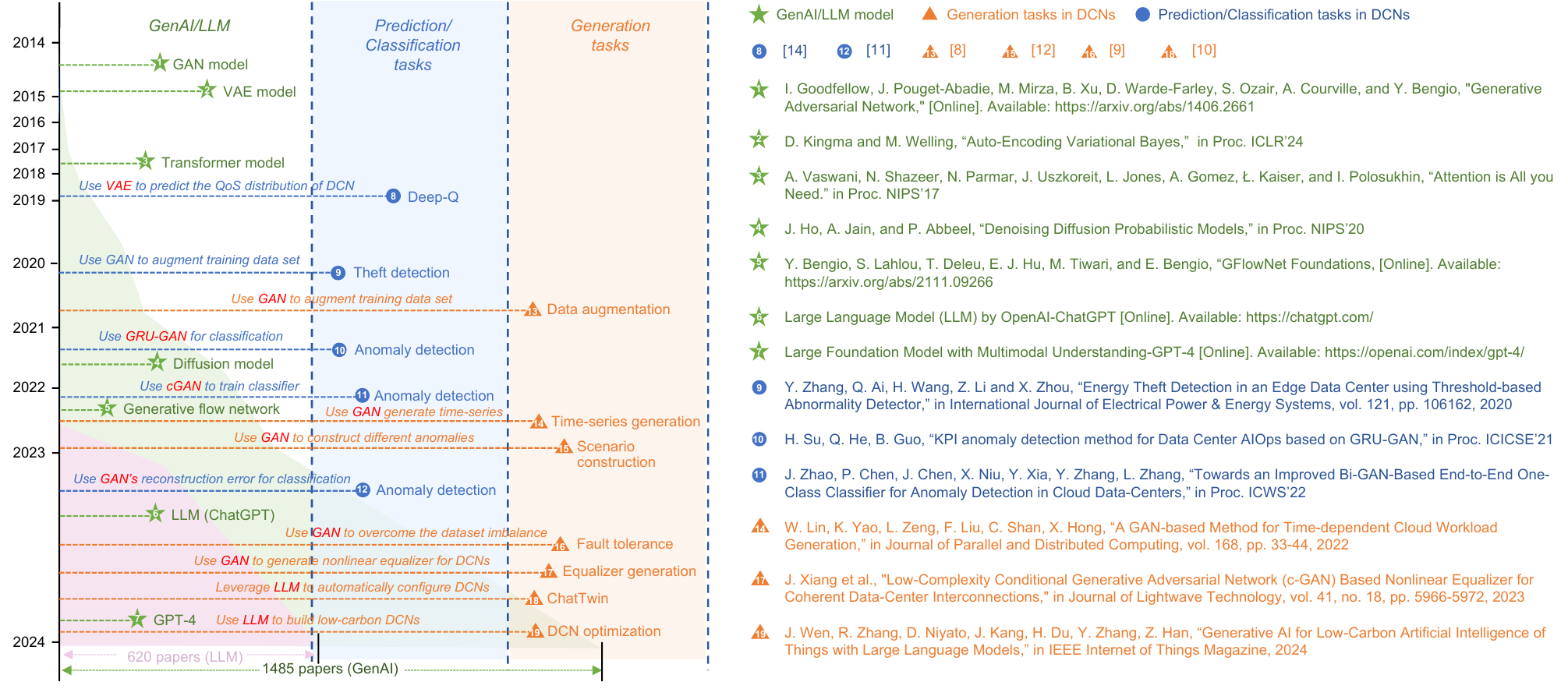} 
  \caption{An illustration of GenAI and LLM's applications in DCNs. The number of peer-reviewed publications regarding GenAI and LLM per year is shown on the left-hand side (the publication data was collected from IEEE Xplore in Sept. 2024).} 
  \label{DCN2}
  \vspace{-0.3cm}
\end{figure*}

\section{Enhancing Data Center Networking with GenAI and LLM}
In this section, we review existing works that apply GenAI and LLMs to enhance DCNs (see Fig. \ref{DCN2}).

\subsection{Data Augmentation}
As a representative generative task, data augmentation in DCNs can be efficiently supported by GenAI.
For instance, the authors in \cite{10.1145/3439602.3439611} presented a GAN-based approach to generate DCN workloads, thereby enriching the datasets to train task scheduling models.
Traditionally, researchers assume that task arrivals in DCNs follow specific distributions (e.g., \textit{Erlang} or \textit{Normal}) and fit models to these assumptions to estimate parameters. 
However, this strategy requires prior knowledge about the workload pattern, which might not always be available or accurate from a practical perspective. 
Moreover, such strategies can only capture simple patterns, whereas current DCN workloads exhibit more complex distributions that are influenced by a variety of unpredictable factors, e.g., instant peak during rush hours or special events. 
Hence, the authors developed a GAN model to synthesize realistic workload data from samples.
The authors evaluated the validity of synthesized data using square error, which is calculated by generating a multi-bin histogram from the generated data and the original data. 
The results show that the GAN-based method can maintain an error of less than 3.558$\times$10$^{-4}$, with much lower time and costs than the conventional methods. 

Similarly, the authors in \cite{DU2023113072} leveraged GANs to address data imbalance in fault diagnosis systems for DCNs, a common challenge due to the rarity of fault instances compared to normal operational data. 
The GAN effectively balanced the training dataset by synthesizing fault data to supplement the limited real fault examples, enhancing the model's ability to recognize and diagnose faults.
The experimental results demonstrate that balanced datasets can reduce the miss alert rate from 10.7\% to 0.5\%.

\subsection{DCN Process Automation}
Building large-scale DCNs is a multifaceted challenge that encompasses server arrangements, network connectivity, cooling systems, and more, all tailored to specific site conditions and services. 
To facilitate the deployment and operation of DCNs, the authors in \cite{10.1145/3600100.3623719} presented ChatTwin, an LLM-empowered conversational system, to generate DCN configurations automatically. 
Specifically, leveraging the advanced natural language understanding capability of ChatGPT-4, ChatTwin first analyzes user requirements through a segment-and-generate approach.
For example, with the user input, \textit{``I need a data hall room with one Air Conditioner Unit (ACU) and eight racks. Each rack contains three servers,"} ChatTwin parses and segments this input into discernible units: one ACU, eight racks, and three servers per rack. 
This segmentation involves identifying key terms and their associated quantities.
Afterward, ChatTwin proceeds to fill out the corresponding DCN configuration files to define each unit. 
Experimental results demonstrate that such an LLM-empowered method can achieve a successful DCN construction rate of 87\% while saving considerable time.

Apart from constructions, a variety of DCN processes can be automated by LLM in a similar way.
For instance, in \cite{NSDI}, the authors employed GPT-4 to automatically compress and analyze DCN operation logs, thus finding implicit anomalies.
The \textit{Uptime Institute Data Center Resiliency Survey 2023}\footnote{The Survey is available at: https://uptimeinstitute.com/resources/research-and-reports/uptime-institute-global-data-center-survey-results-2023} reveals that 39\% of data center operators have experienced a serious outage because of human error, of which 50\% were the result of a failure to follow the correct procedures. 
In this case, leveraging LLMs or foundation models can significantly mitigate these issues by automating procedural compliance and decision-making processes in DCNs. 
Additionally, LLMs can train personnel by providing real-time feedback, reducing the risk of procedural deviations that lead to serious outages. 


\subsection{Anomaly Detection}
Given the stringent requirements of DCNs for reliability, anomalies should be detected and tackled promptly.
For instance, sudden, unexpected increases in network traffic that deviate significantly from normal patterns might be indicative of denial-of-service attacks.
GenAI has been widely adopted for anomaly detection in the DCN context.
The authors in \cite{AD} proposed an innovative approach to DCN anomaly detection via autoencoders.
The autoencoder is trained with only normal traffic data and is expected to recover any given input as close as possible to the learned normal patterns. 
Therefore, an input instance can be classified as an attack if its reconstruction error is larger than a predefined threshold; otherwise, it is regarded as normal.
We can observe that the strong distribution representation capability enables GenAI models to support not only generation but also classification tasks.
Furthermore, GenAI's adaptability is showcased through its ability to construct diverse conditions, thus benchmarking the robustness of DCNs facing various anomalies. 
For instance, the authors in \cite{Anomaly} developed a conditional GAN, utilizing a Gaussian noise with adjustable standard deviation to indicate how anomalies are desired in the generated traffic patterns.
Afterward, various conditions can be synthesized, which ensures that the DCN can be rigorously tested against a spectrum of potential issues.

\subsection{Quality of Service Modeling}
Apart from classification, GenAI models can facilitate predictive tasks in DCN.
The authors in \cite{10.1145/3229543.3229549} developed a deep generative network for DCNs named Deep-Q, leveraging a VAE to predict the QoS under different traffic load conditions.
Intuitively, DAI models can solve this problem by learning the mapping relationship between input and output and predicting the exact QoS value under the given traffic load. 
In contrast, GenAI generates QoS distributions, which more accurately reflect DCN realities, as many unseen factors, such as packet loss and queue dynamics, make QoS measures follow a complex distribution rather than deterministic. 
Experimental results proved that Deep-Q achieves 3$\times$ higher QoS prediction accuracy than traditional methods. 

\textbf{Perspectives}: Our review illustrates that GenAI can not only independently address DCN tasks such as process automation, but also effectively enhance existing DAI methods. As shown in Fig. \ref{DCN}, GenAI can expand DAI training data, help DAI extract information from complex environments, dynamically select DAI models, etc. These collaborative paradigms need further research.

\section{Transforming DCNs for GenAI/LLM}
In this section, we present how DCNs evolve to efficiently support GenAI and LLM.
\begin{figure*}[tbp!]
  \centering
  \includegraphics[width=0.95\textwidth]{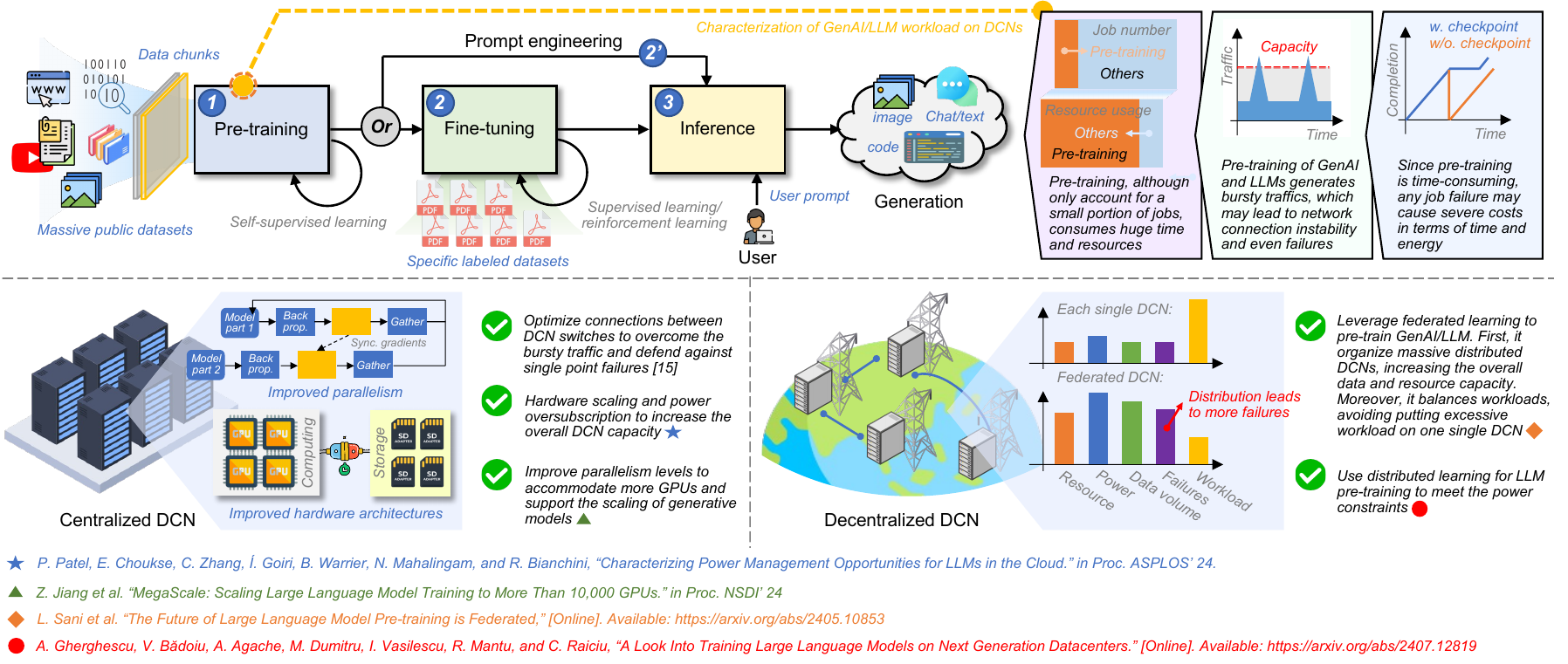} 
  \caption{Top: The illustration of GenAI/LLM lifecycle. We can observe that most of the issues happen in the pre-training stage. Bottom: The centralized and decentralized DCNs.} 
  \label{DCN3}
  \vspace{-0.3cm}
\end{figure*}

\subsection{Characterization of GenAI/LLM Workload}
GenAI/LLM in DCNs follows a distinct lifecycle with three key stages. 
As shown in Fig. \ref{DCN3}(top), pre-training is initially performed, where GenAI models are trained on expansive datasets in a self-supervised manner to accumulate generalizable knowledge and construct a broad capability to handle diverse tasks (e.g., conversation and text-to-image generation). 
Afterward, the GenAI models undergo task-oriented fine-tuning, usually trained on specific labeled datasets via supervised or reinforcement learning. 
This fine-tuning will align the GenAI models with the particular requirements of certain downstream tasks. 
Note that such alignment can also be realized by prompt engineering, i.e., users craft the prompts for instructing GenAI/LLM during the inference stage.
Finally, well-trained models are deployed to perform GenAI inferences and accomplish user demands. 

According to a six-month trace of LLM workload in practical DCNs, the authors in \cite{NSDI} stated that the discrepancies between GenAI/LLMs and prior task-specific deep learning workload exist in the following three aspects.
\begin{itemize}
    \item \textbf{Imbalanced Resource Usage}: The resource usage of LLMs exhibits a notable imbalance. As shown in Fig. \ref{DCN3}(top), pre-training tasks, though they represent merely 3.2\% of the total job count, disproportionately consume 94.0\% of all GPU resources. In contrast, generative inferences, which make up 92.9\% of job activities, utilize only 0.8\% of the available resources \cite{NSDI}. Secondly, there is an uneven allocation among the infrastructure components. Specifically, while CPU, host memory, and network resources are often underutilized, GPU resources are consistently at high utilization levels \cite{NSDI}. 
    \item \textbf{Bursty Traffic Patterns}: Traditional cloud computing tasks typically handle millions of flows with traffic utilization averaging below 20\%, exhibiting a steady and gradual pattern that changes on an hourly basis \cite{SIGCOMM}. In contrast, LLM training involves infrequent but intense bursty flows. The Network Interface Card (NIC) sporadically handles substantial data volumes, quickly reaching full network capacity and sustaining this level for several seconds to tens of seconds \cite{SIGCOMM}. This pattern stems from the necessity for gradient synchronization during LLM training. Each training iteration requires synchronization of data across multiple GPU groups during the backward phase, resulting in sharp traffic spikes.
    \item \textbf{Sensitive to Failures}: According to \cite{NSDI}, nearly 40\% LLM pre-training jobs fail, mainly caused by infrastructure failures, GPU overheating, and connection errors \cite{NSDI, SIGCOMM}. In contrast, inference jobs seldom encounter failures since they only call the well-trained generative models. Given the long period of pre-training, failure recovery is extremely costly. Nowadays, the GenAI/LLM training typically leverages checkpoints to recover from failures. Nonetheless, generating a checkpoint requires substantial storage (e.g., 30GB per GPU in \cite{SIGCOMM}) and incurs high overhead (e.g., 100s in \cite{SIGCOMM}).
\end{itemize}

\subsection{DCNs in the GenAI/LLM Era}
Facing the unique workload patterns of GenAI/LLM in DCNs, researchers are exploring mechanisms to optimize data center architectures, ensuring more efficient and robust operations. 
First, federated learning can be leveraged to overcome the tremendous resource overhead of GenAI/LLM pre-training.
By organizing geographically distributed DCNs and scheduling their computing and data resources, the pre-training workload can be balanced, avoiding assigning excessive burdens to one certain server. 
To enhance the robustness of DCNs against bursty traffics \cite{SIGCOMM}, the authors presented the dual Top-of-Rack (ToR) architecture, which significantly reduces failure rates of GenAI/LLM pre-training tasks by eliminating dependencies on direct links between switches.
Furthermore, the authors in \cite{NSDI} presented fault-tolerant pretraining solutions that enhance reliability through asynchronous checkpoints and LLM-assisted automated diagnosis. 

\begin{figure*}[htbp]
  \centering
  \includegraphics[width=0.9\textwidth]{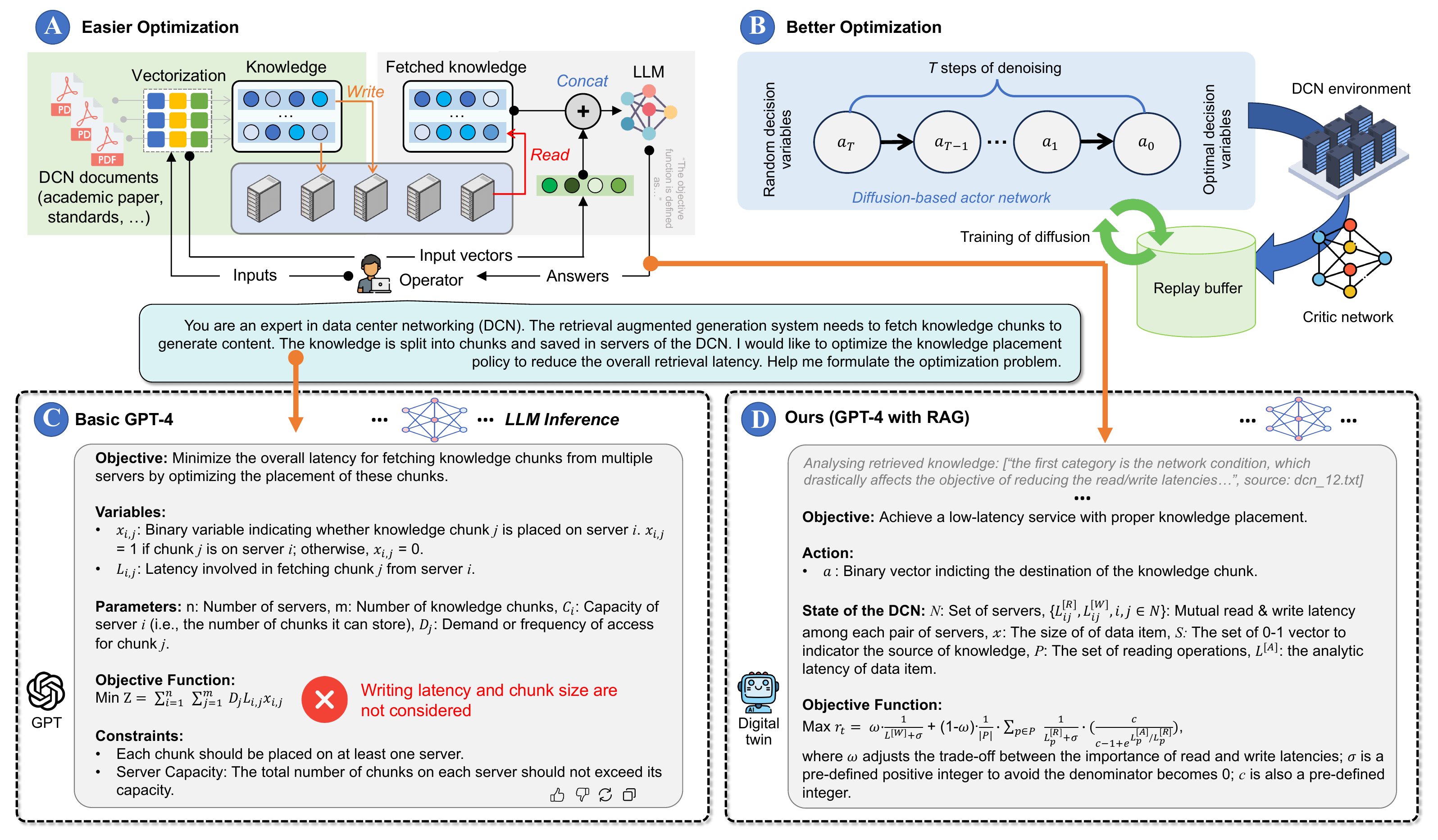} 
  \caption{A: The automatic optimization formulation. B: Diffusion-DRL for optimization solving. C \& D: The comparison of optimization formulation between the backbone GPT-4 with the proposed digital twin.} 
  \label{DCN4}
  \vspace{-0.4cm}
\end{figure*}

Apart from building efficient GenAI/LLM models, the sustainability of DCNs also attracts great attention, due to the considerable resource consumption of pre-training.
For instance, POLCA is a power oversubscription framework that capitalizes on the unused power headroom by allowing 30\% more server deployments within existing power infrastructures. 
From the task perspective, hybrid DCN leverages a cost-based scheduling framework to dynamically assign LLM tasks to the most suitable hardware. 
It optimizes task allocation by deciding whether tasks should be processed on energy-efficient processors or high-performance GPUs, guided by the size of each task's input and output tokens. 
Such a workload-aware approach decreases GPU energy consumption by 7.5\% compared to a workload-unaware baseline.
Furthermore, cloud-edge collaboration is proposed to realize energy-efficient GenAI/LLM. 
By integrating edge databases, the generated content/response can be cached, significantly reducing the frequency and volume of data queries needing cloud processing.  

\textbf{Perspectives}: Looking ahead, we believe that DCNs will evolve in two directions (see Fig. \ref{DCN3}(bottom)). First, centralized DCNs will continue to optimize storage, communication, and computing efficiency (such as by reducing transmission and memory read/write latency) to improve parallelism and performance, thereby supporting large generative models. Meanwhile, decentralized DCNs based on distributed or federated learning will be further developed to integrate massive resources and data over the network.

\section{Case Study: Towards Full-Lifecycle DCN Digital Twin via GenAI}
In this section, we perform a case study using LLM to implement the digital twin that covers the entire DCN lifecycle.

\subsection{Problem Statement}
From the operator's perspective, the complete DCN lifecycle consists of configuration, optimization, and management phases. 
First, configuration refers to designing DCNs, setting the number of servers, the layout of racks, and connections \cite{10.1145/3600100.3623719}. 
Optimization entails improving strategies for routing, resource allocation, load balancing, etc. 
Finally, management focuses on ensuring the DCN operations, including anomaly detection, fault diagnosis, and recovery \cite{NSDI}. 
\textit{Consequently, a full-lifecycle digital twin should realize the end-to-end automation of all these stages.}
In \cite{10.1145/3600100.3623719} and \cite{NSDI}, researchers demonstrated the effectiveness of LLMs in designing and managing DCNs, respectively. 
Hence, this case study addresses the research gap toward full-lifecycle DCN digital twins by exploring automatic DCN optimization.

\subsection{Proposed Framework}
Our framework consists of two modules, intending to make DCN optimization easier and better, respectively.
\subsubsection{Automatic Optimization Formulation}
DCN operators need to formulate optimization problems manually, which involves gathering the relevant factors and modeling their mutual effects on the objective via appropriate formulas.
However, this process is time-consuming and heavily relies on the operator's experience.
Moreover, given that DCN environments exhibit great heterogeneity, human-crafted formulations may fail to reflect the complicated but implicit mutual effects of factors precisely.
To this end, we present the automatic DCN optimization formulation leveraging an LLM with RAG.

As shown in Fig. \ref{DCN4}-A, the LLM is first fed with DCN-related documents (academic papers, standards, etc.), thereby increasing its expertise in DCN-related optimizations, such as load-balancing and routing.
Specifically, textual documents are vectorized into word embeddings, which are split into chunks and saved in the DCN.
Then, operators can formulate the optimization problem in a conversational way.
In detail, the operator's description of the DCN state, optimization objective, and key factors are vectorized.
Afterward, the digital twin calculates the semantic distance between inputs and all preserved chunks.
The most relevant chunks will be fetched and serve as external knowledge, with which LLM is further employed to generate coherent responses that promote the process of problem formulation.

\subsubsection{Diffusion-empowered Optimization Solving}
To solve the formulated optimization problems, we leverage the DRL paradigm and adopt Diffusion-DRL \cite{10515203}.
Particularly, it follows the actor-citric architecture while building the actor network via diffusion.
As shown in Fig. \ref{DCN4}-B, in each training round, the decision variables are initialized randomly and sent to the denoising network.
Following the diffusion principle, the noisy decision variables are de-noised through a Markov denoising process.
The final decision variables are then put into the specific DCN environment and acquire the rewards, which are designed as the Q-value and guide the refinement of diffusion's parameters.
The state, action, and reward are defined as follows.
\begin{itemize}
    \item \textbf{State}: State describes the DCN environment, including the values of all the factors that affect the reward calculation. 
    \item \textbf{Action}: The decision variable that should be optimized.
    \item \textbf{Reward}: The efficiency of applying the learned action in the current state, indicating the desirability of the actions.
\end{itemize}

\subsection{Evaluation}
To evaluate the effectiveness of the proposed digital twin, we first leverage it to model a representative DCN optimization problem, i.e., data placement.
Recall that each DCN contains numerous interconnected servers.
The server selection when writing data to DCN significantly affects the service latency (e.g., the RAG latency).
To this end, we send baseline GPT-4 and our digital twin with the same prompt, which describes the objective of minimizing RAG latency by optimizing the placement of knowledge chunks.
As shown in Fig. \ref{DCN4}-C, the baseline GPT-4's output is general and superficial.
Although it successfully models the objective of latency minimization, various factors that may affect the performance are ignored.
In contrast, contributed to RAG, our digital twin routes to the most relevant documents via semantic router (see Fig. \ref{DCN4}-D).
Then, it fetches the key expertise, fuses it with the pre-trained knowledge of LLMs, and generates the final problem formulation.
We can observe that both the knowledge writing and the following reading operations are considered.
Moreover, the objective is defined as the weighted sum of the reciprocals of read \& write latencies, thus reflecting the pursuit of minimizing RAG latency.

\begin{figure}[tbp!]
  \centering
  \includegraphics[width=0.5\textwidth]{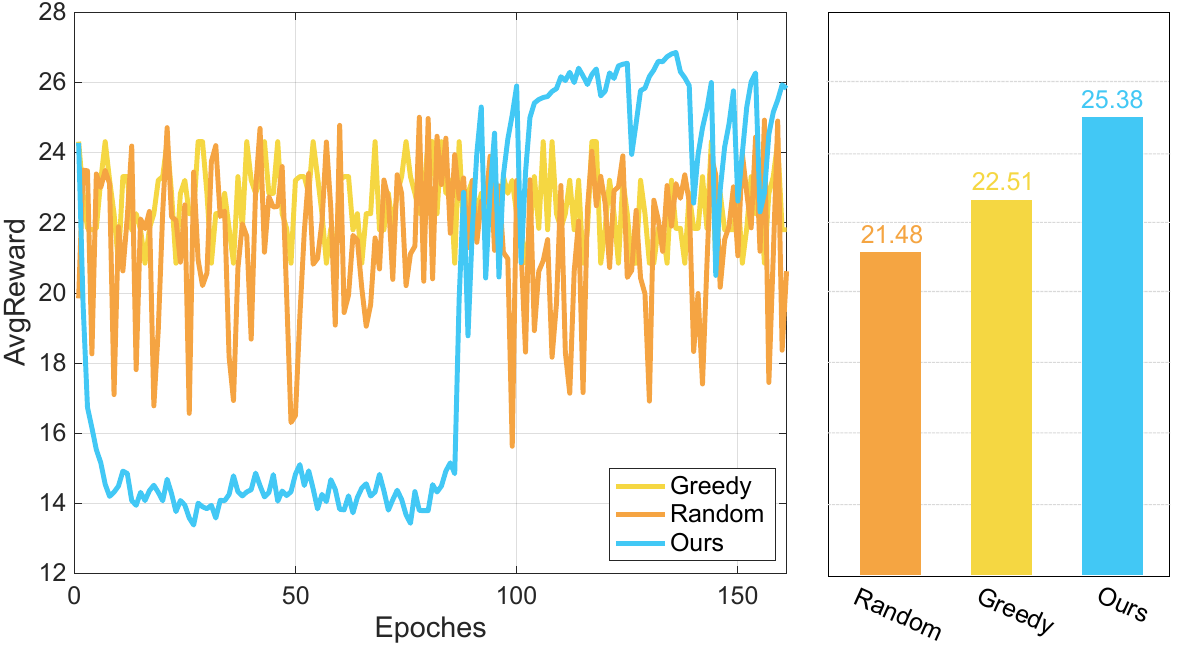} 
  \caption{The training curve and rewards of random, greedy, and the proposal knowledge placement policies.} 
  \label{DCN6}
\end{figure}

Then, we explore the efficiency of the adopted Diffusion-DRL in solving optimization problems.
Particularly, we implement two default strategies, i.e., the knowledge is randomly placed (named \textit{Random}), and the knowledge is placed in the serve with the lowest reading latency (named \textit{Greedy}).
As shown in Fig. \ref{DCN6}, the random policy leads to dramatic performance fluctuation, which may cause stragglers.
Aiming at minimizing the reading latency, the greedy policy achieves higher rewards and a higher level of stability.
Finally, the proposed Diffusion-DRL approach outperforms the baselines and can efficiently reduce the retrieval latency, thus facilitating the DCN digital twin to manage large-scale knowledge.

\section{Future Research Directions}
In this section, we outline three main future directions for deepening the interplay between GenAI/LLM and DCN.
\begin{itemize}
    \item \textbf{LLM-enhanced DCN Optimization}: Apart from diffusion models that enhance the exploration capability of DRL, LLM also facilitates DCN decision-making. For instance, LLM can serve as automatic agents and provide human-like feedback to each action, thus improving the human-centric DCN performance.
    \item \textbf{Embodied DCN Digital Twin}: LLM is the building block for constructing embodied DCN digital twins. First, its multimodal understanding capability enables the digital twin to perceive complex physical environments. Moreover, LLMs' large size allows life-long continuous learning, allowing detail twins to keep reinforcing their policies by fine-tuning.
    \item \textbf{DCN Security in the LLM Era}: Future research should focus on developing robust mechanisms to safeguard against adversarial attacks tailored to exploit the vulnerabilities of LLM-driven systems, such as the poisoning attacks toward training data and deepfake.
\end{itemize}

\section{Conclusion}
This article has explored the interplay between GenAI and DCN.
From the perspective of how GenAI/LLM enhances DCNs, we have introduced the existing challenges in the DCN field and reviewed the literature about how GenAI/LLM offers new possibilities.
Then, shifting the viewpoint to how DCN evolved to serve GenAI/LLM, we analyzed the characteristics of GenAI/LLM workloads on DCNs as well as the corresponding requirements.
Finally, we have performed a case study on developing full-lifecycle DCN digital twins.

\bibliographystyle{IEEEtran}
\bibliography{dcn}
\vfill

\end{document}